\documentstyle[epsf,twoside,fleqn,espcrc2]{article}

\newcommand{\AmS}{{\protect\the\textfont2
   A\kern-.1667em\lower.5ex\hbox{M}\kern-.125emS}}

\title{Exclusive Radiative Decays of B Mesons\thanks{Invited talk at the 9th International Conference in Quantum Chromodynamics (Montpellier 2--9th July 2002)}}

\author{Stefan W. Bosch\address{Max-Planck-Institut f\"ur Physik,
 Werner-Heisenberg-Institut, \\F\"ohringer Ring 6, D-80805 Munich, Germany}}

\begin{document}

\begin{abstract}
We present within the Standard Model the exclusive radiative decays $B\to K^* /\rho \gamma$ and $B_{s/d}\to\gamma\gamma$ in QCD factorization based on the heavy-quark limit $m_b\gg\Lambda_{QCD}$. For the decays with a vector meson in the final state we give results complete to next-to-leading order in QCD.
\end{abstract}

\maketitle

\section{Introduction}

The main goal of todays $B$ physics is a precision study of the flavour sector and the phenomenon of CP violation that comes along with it. Particularly useful for clean tests of the Standard Model are the rare radiative $b\to s (d)\gamma$ transitions. The inclusive $b\to s\gamma$ mode was measured to have a branching fraction of $B(B\to X_s\gamma)=(3.23\pm 0.42)\cdot 10^{-4}$. Branching ratios of exclusive radiative channels are measured to be $B(B^0\!\to K^{*0}\gamma)=(4.44\pm 0.35)\cdot 10^{-5}$ and $B(B^+\!\to K^{*+}\gamma)=(3.82\pm 0.47)\cdot 10^{-5}$ whereas for the $B\to\rho\gamma$ and $B\to\gamma\gamma$ decays so far only upper limits exist \cite{meas}.

Whereas the inclusive mode can be computed perturbatively using the heavy-quark expansion, for the exclusive channels bound state effects have to be taken into account. The basic mechanisms for the exclusive radiative decays were already discussed by various groups \cite{excl}. However, they all had to use hadronic models which do not allow a clear separation of short- and long-distance dynamics and a clean distinction of model-dependent and model-independent features.

Yet, a systematic and model-independent analysis of exclusive radiative decays is possible in the heavy quark limit $m_b\gg\Lambda_{QCD}$. The relevant hadronic matrix elements of local operators in the weak Hamiltonian simplify in this limit because perturbatively calculable hard scattering kernels can be separated from nonperturbative form factors and universal light-cone distribution amplitudes. A power counting in $\Lambda_{QCD}/m_b$ implies a hierarchy among the possible transition mechanisms and allows to identify leading and subleading contributions. In particular, effects from quark loops are calculable rather than being generic, uncalculable long-distance contributions. Our approach is similar in spirit to the treatment of hadronic matrix elements in two-body non-leptonic $B$ decays formulated by Beneke, Buchalla, Neubert, and Sachrajda \cite{BBNS}.

\section{$B\to V\gamma$ at NLO in QCD}
The effective Hamiltonian for $b\to s\gamma$ transitions reads
\begin{equation}\label{heff}
  {\cal H}_{eff}=\frac{G_F}{\sqrt{2}}\sum_{p=u,c}\lambda_p^{(s)}
\bigg( \sum_{i=1}^2 C_i Q^p_i +\sum_{j=3}^8 C_j Q_j\bigg)
\end{equation}
where $\lambda_p^{(s)}=V^*_{ps}V_{pb}$. The relevant operators are the current-current operators $Q^p_{1,2}$, the QCD-penguin operators $Q_{3\ldots 6}$, and the electro- and chromomagnetic penguin operators $Q_{7,8}$. The effective Hamiltonian for $b\to d\gamma$ is obtained from (\ref{heff}) by the replacement $s\to d$. The most difficult step in computing the $B\to V\gamma$ decay amplitudes is the evaluation of the hadronic matrix elements of the operators in (\ref{heff}). In the heavy-quark limit a systematic treatment of the hadronic matrix elements is possible. In this case the following factorization formula \cite{BVgam} is valid
\begin{eqnarray}\label{fform}
  \lefteqn{\langle V\gamma(\epsilon)|Q_i|\bar B\rangle =} \nonumber\\
  &&\hspace*{-4ex} =\Big[ F^{B\to V} T^I_{i} + \!\int^1_0 \!\!d\xi\, dv \, T^{II}_i(\xi,v) \Phi_B(\xi) \Phi_V(v)\Big] \!\cdot\epsilon \nonumber
\end{eqnarray}
where $\epsilon$ is the photon polarization 4-vector. Here $F^{B\to V}$ is a $B\to V$ transition form factor, and $\Phi_B$, $\Phi_V$ are leading twist light-cone distribution amplitudes of the $B$ meson and the vector meson $V$, respectively. These quantities are universal, nonperturbative objects. They describe the long-distance dynamics of the matrix elements, which is factorized from the perturbative, short-distance interactions expressed in the hard-scattering kernels $T^I_{i}$ and $T^{II}_i$. To leading order in QCD and leading power in the heavy-quark limit, $Q_7$ gives the only contribution to the $B\to V\gamma$ amplitude. At ${\cal O}(\alpha_s)$ the operators $Q_{1\ldots 6}$ and $Q_8$ start contributing and the factorization formula becomes nontrivial.

The relevant diagrams for the NLO hard-vertex corrections $T_i^I$ were computed in \cite{GHWBCMU} to get the virtual corrections to the matrix elements for the inclusive $b\to s\gamma$ mode at next-to-leading order. We re-interpret these results as the perturbative type I hard-scattering kernels for the exclusive process. As required for the consistency of the factorization formula, these contributions are dominated by hard scales $\sim m_b$ and are hence infrared finite.

\begin{figure}
\epsfxsize=7.5truecm
\centerline{\epsffile{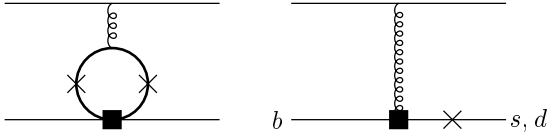}}
\vspace{-0.8truecm}
\caption{\label{fig:qit2} ${\cal O}(\alpha_s)$ and leading power contribution to the hard-scattering kernels $T^{II}_i$ from four-quark operators $Q_i$ (left) and from $Q_8$. The crosses indicate the places where the emitted photon can be attached.}
\vspace{-0.5truecm}
\end{figure}

We now turn to the mechanism where the spectator participates in the hard scattering. The non-vanishing contributions to $T^{II}_i$ are shown in Fig.~\ref{fig:qit2}. 
We can express both the type I and type II contributions to the matrix elements $\langle Q_i\rangle$ in terms of the matrix element $\langle Q_7\rangle$, an explicit factor $\alpha_s$, and hard-scattering functions $G_i$ and $H_i$ which are given explicitely in \cite{thesis}.\bigskip

\begin{figure}[t]
\hbox{\epsfxsize=3.75truecm\epsffile{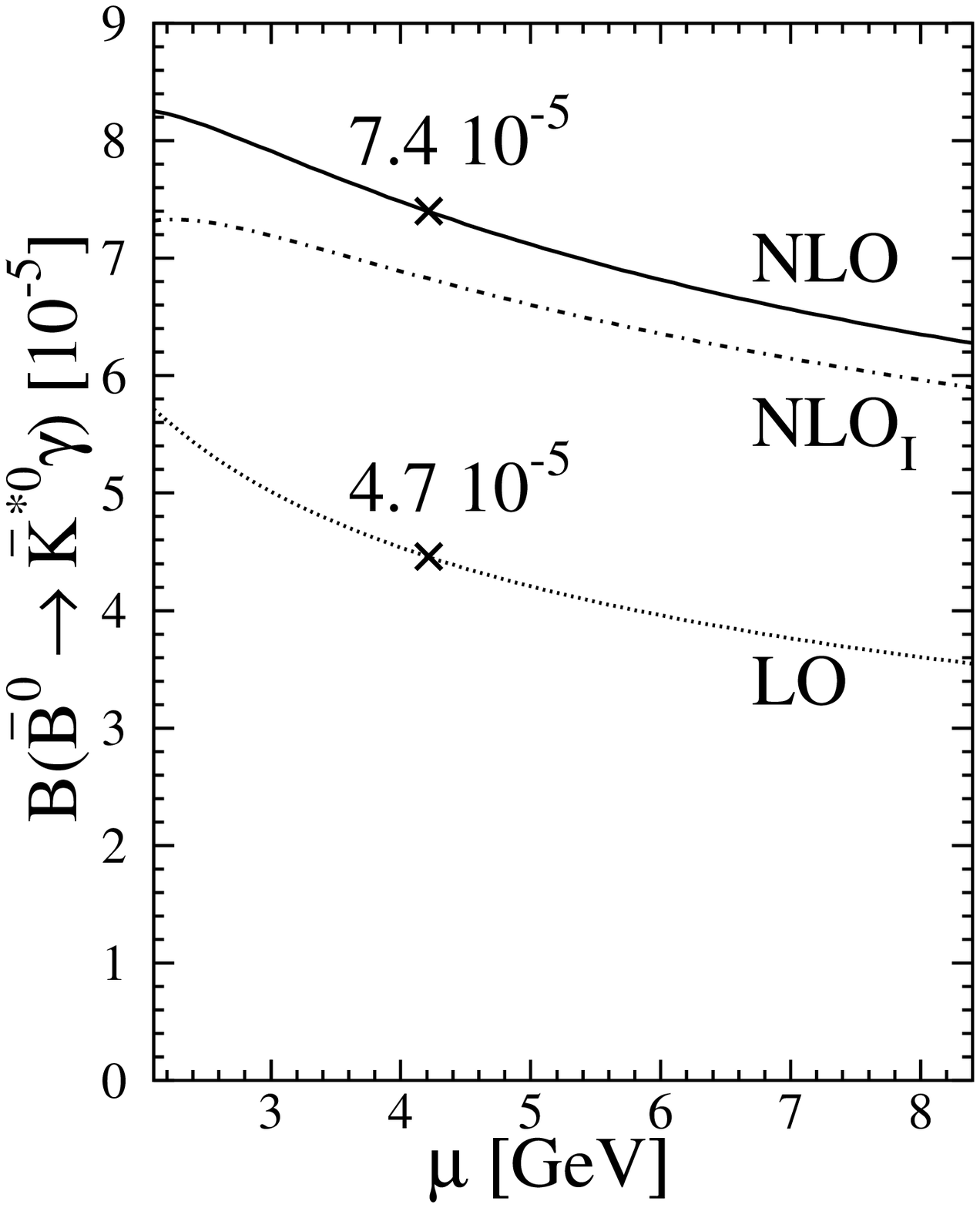}\epsfxsize=3.75truecm\epsffile{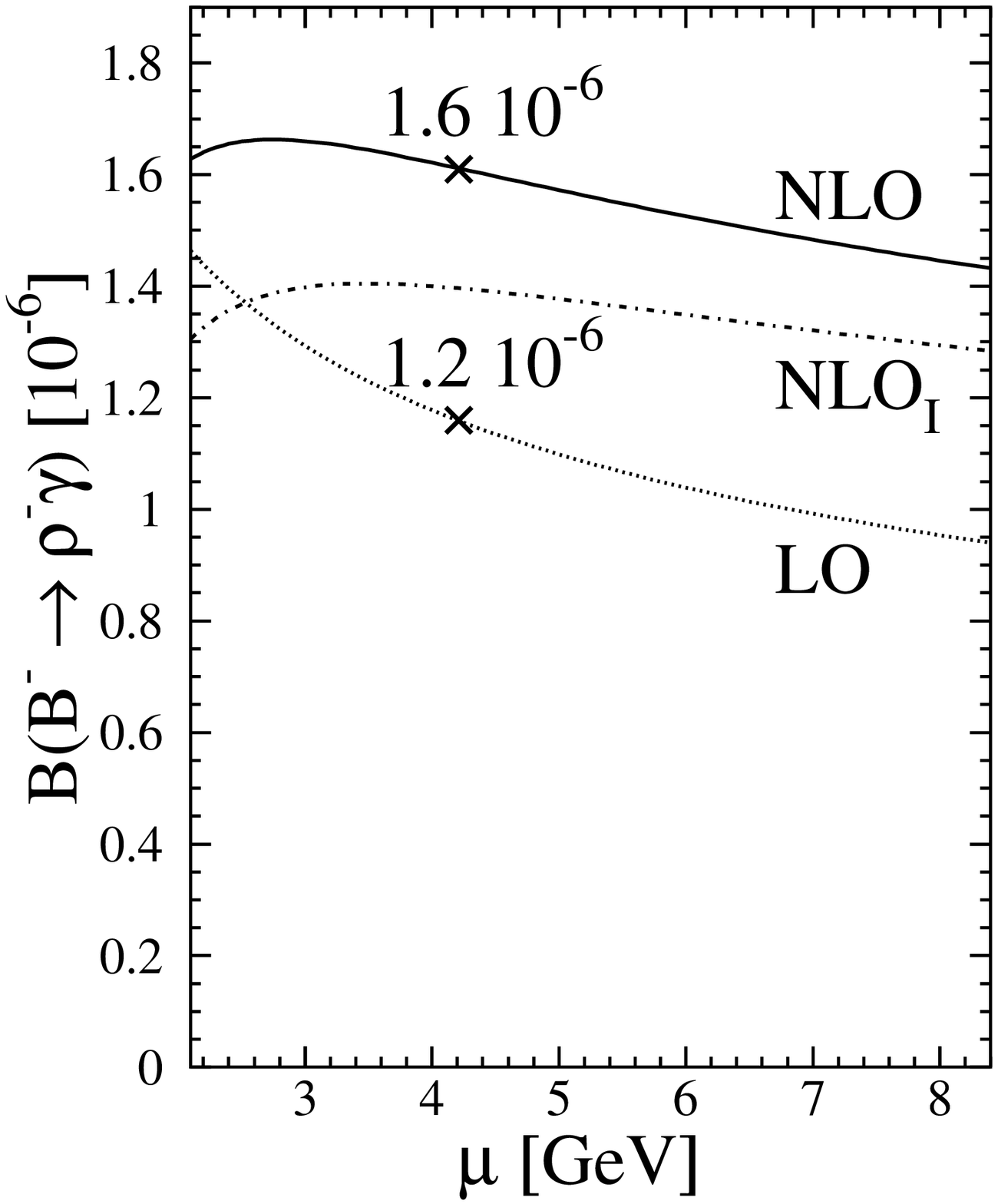}}
\vspace{-0.8truecm}
\caption{Dependence of the branching fractions $B(\bar{B}^0 \to \bar{K}^{*0} \gamma)$ and $B(B^- \to \rho^- \gamma)$ on the renormalization scale $\mu$. The dotted line shows the LO, the dash-dotted line the NLO result including type-I corrections only and the solid line shows the complete NLO result.}
\label{fig:bkrhomu}
\vspace{-0.5truecm}
\end{figure}

The total $\bar B\to V\gamma$ amplitude then can be written as
\[A(\bar B\to V\gamma)=\frac{G_F}{\sqrt{2}}\left[\lambda_u^{(s)}a^u_7 +\lambda_c^{(s)} a^c_7\right]\langle V\gamma|Q_7|\bar B\rangle\]
where the factorization coefficients $a_7^p(V\gamma)$ consist of the Wilson coefficient $C_7$ and the contributions from the type I and type II hard-scattering corrections. We get a sizeable enhancement of the leading order value, dominated by the $T^I$-type correction.
The net enhancement of $a_7$ at NLO leads to a corresponding enhancement of the branching ratios, for fixed value of the form factor. This is illustrated in Fig. \ref{fig:bkrhomu},
where we show the residual scale dependence for $B(\bar{B}\to \bar{K}^{*0}\gamma)$ and $B(B^-\to\rho^-\gamma)$ at leading and next-to-leading order.

A complex phase is generated at NLO which leads to a non-vanishing value of the CP asymmetry
\[{\cal A}_{CP}(V\gamma)=\frac{\Gamma(B\to V\gamma)-\Gamma(\bar B\to V\gamma)}{\Gamma(B\to V\gamma)+\Gamma(\bar B\to V\gamma)}\]
It is of ${\cal O}(10\%)$ for the $\rho\gamma$ mode with the largest theoretical uncertainty coming from the scale dependence. The CP asymmetry is with ${\cal A}_{CP}(K^*\gamma)\approx -0.3\%$ very small for the $b\to s\gamma$ transition because of the large CKM hierarchy $|\lambda_u^{(s)}| \ll |\lambda_c^{(s)}|$.

\begin{figure}[t]
\hbox{\epsfxsize=3.75truecm\epsffile{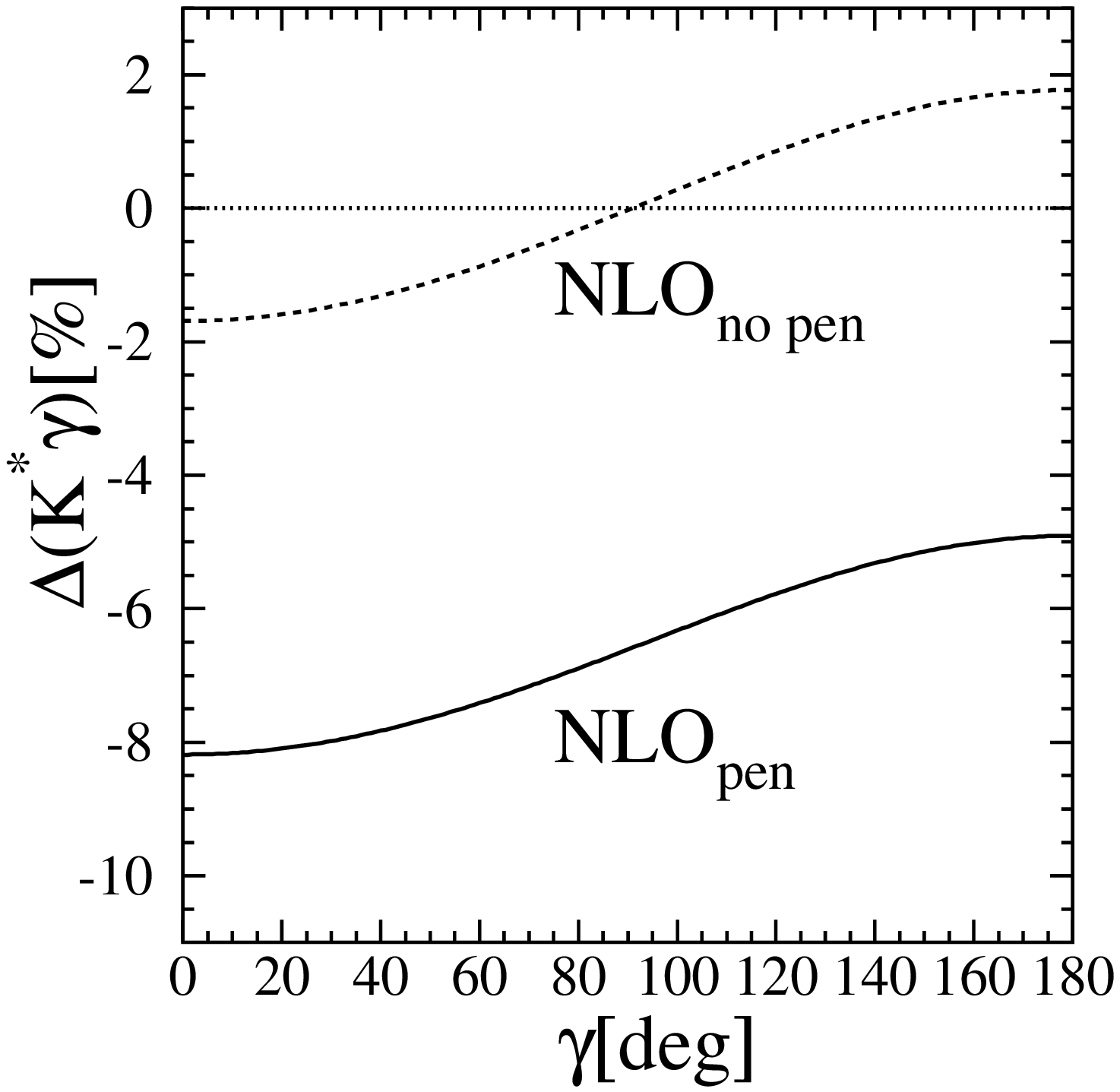}\epsfxsize=3.75truecm\epsffile{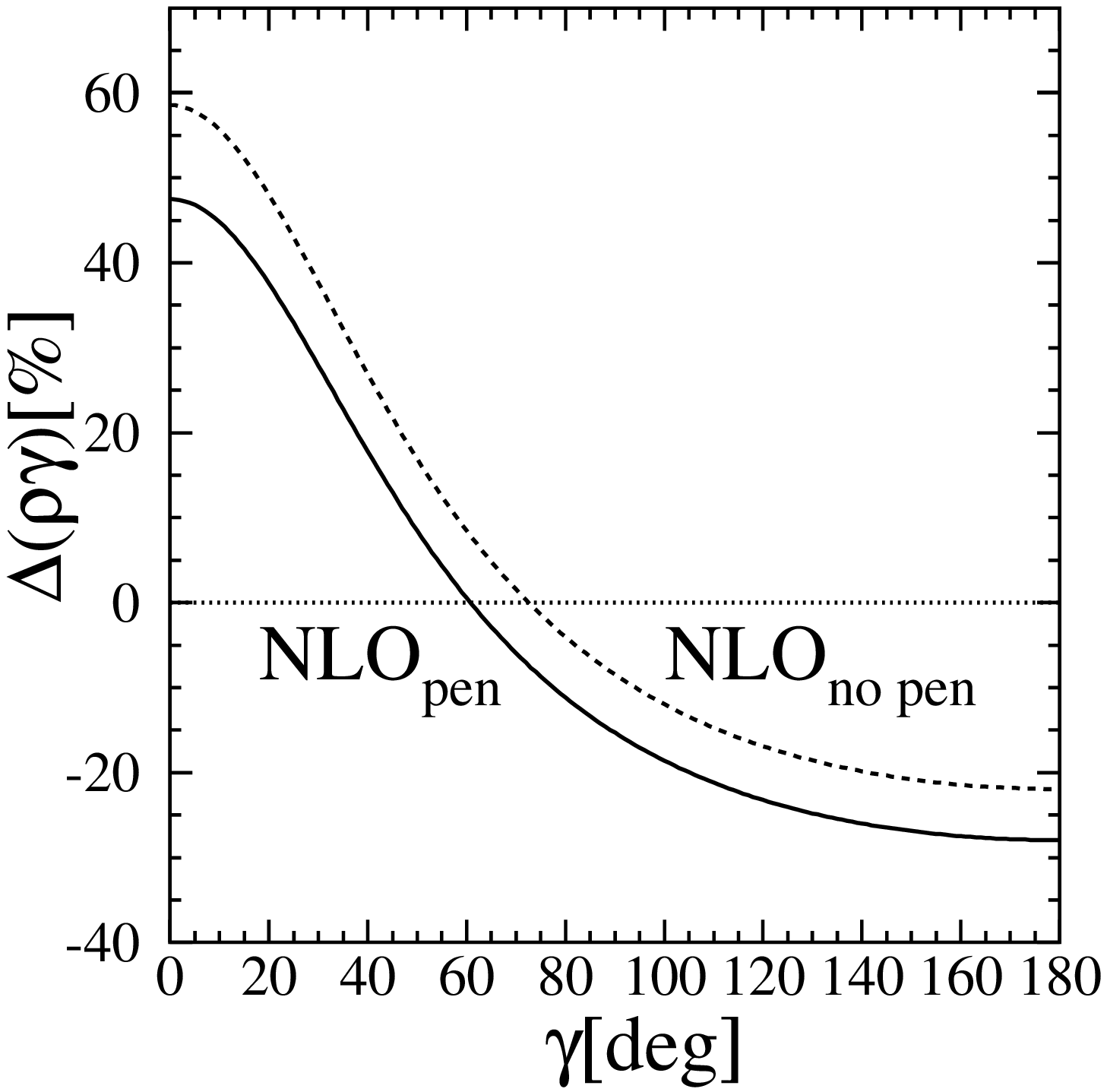}}
\vspace{-0.8truecm}
\caption{The isospin-breaking asymmetries $\Delta(K^*\gamma)$ and $\Delta(\rho\gamma)$ as a function of the CKM angle $\gamma$ with and without the inclusion of QCD penguin operator effects.}
\label{fig:isodelta}
\vspace{-0.5truecm}
\end{figure}

A further interesting observable is the charge averaged isospin breaking ratio
\[\Delta(V\gamma)=\frac{v\,\Gamma(B^+\!\!\to V^+\gamma)}{2\Gamma(B^0\!\to V^0\gamma)} \!\!\;+\!\!\; \frac{v\,\Gamma(B^-\!\!\to V^-\gamma)}{2\Gamma(\bar{B}^0\!\to V^0\gamma)}-1\]
where $v\!=\!1$ for $V\!=\!K^*$ and $v\!=\!1/2$ for $V\!=\!\rho$. Within our approximations, isospin breaking is generated by weak annihilation contributions. They are power suppressed but can nevertheless be computed within QCD factorization because the colour-transparency argument applies to the emitted, highly energetic vector meson in the heavy-quark limit. Isospin breaking was already discussed in \cite{AHL}, partially including NLO corrections. Kagan and Neubert found a large effect from the penguin operator $Q_6$ on the isospin asymmetry $\Delta (K^*\gamma)$ \cite{KN}. This brings the prediction $\Delta(K^*\gamma)=(-7.5^{+4.1}_{-5.9})\%$ within the large error bars in rather good agreement with the experimental value $\Delta(K^*\gamma)_{exp}=(-19.2\pm 11.8)\%$ \cite{thesis}. For $B\to\rho\gamma$ we find a strong dependence of the isospin asymmetry on the angle $\gamma$ of the unitarity triangle. This is illustrated in Fig.~\ref{fig:isodelta}.
A measurement of the isospin asymmetry $\Delta(\rho\gamma)$ can therefore give valuable indirect information on the CKM angle $\gamma$.

\section{The Radiative Decays $B\to\gamma\gamma$}
The double radiative $B\to\gamma\gamma$ modes realize the exceptional situation of nontrivial QCD dynamics related to the decaying $B$ in combination with a completely nonhadronic final state and simple two-body kinematics. As the two-photon system can be in a CP-even 
or CP-odd state 
we can study direct CP-violating effects.

In the heavy-quark limit we get a factorization formula \cite{thesis,BBgamgam} for the hadronic matrix elements of the operators in the effective Hamiltonian (\ref{heff}):
\[ \langle \gamma(\epsilon_1)\gamma(\epsilon_2)|Q_i|\bar B\rangle = \int^1_0 d\xi\, T^{\mu\nu}_i(\xi)\, \Phi_{B1}(\xi)\epsilon_{1\mu} \epsilon_{2\nu}\]
Because there are no hadrons in the final state only one type of hard-scattering kernel $T$ (type II or hard-spectator contribution) enters the factorization formula. 

To leading power in $\Lambda_{QCD}/m_b$ and in the leading logarithmic approximation of QCD only one diagram contributes to the amplitude for $B\to\gamma\gamma$. It is the one-particle reducible (1PR) diagram with the electromagnetic penguin operator $Q_7$, where the second photon is emitted from the $s$-quark line. The result for both the CP-even and CP-odd amplitude is
\[A_\pm = -(\lambda^{(q)}_u +\lambda^{(q)}_c) C_7 \frac{m_{B}}{\lambda_{B}}\]
where $\lambda_B={\cal O}(\Lambda_{QCD})$ parametrizes the first negative moment of the $B$ meson wave function. In the present approximation the strong-interaction matrix elements multiplying $\lambda^{(q)}_u$ and $\lambda^{(q)}_c$ are identical and have no relative phase. For the strictly leading-power result we therefore get no direct CP violation.

\begin{figure}[t]
\centerline{\hbox{\epsfxsize=4.3truecm\epsffile{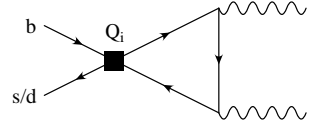}}}
\vspace{-0.8truecm}
\caption{The subleading power 1PI diagram.\label{fig:1PI}}
\vspace{-0.5truecm}
\end{figure}

Subleading contributions come for example from the 1PR diagram, where the second photon is emitted from the $b$ quark line. This and other subleading effects in $\langle Q_7\rangle$, however, contribute equally to the up- and charm-quark components of the amplitude and therefore do not affect direct CP violation. The one-particle irreducible diagrams (1PI) of Fig. \ref{fig:1PI},
on the other hand, provide the basic effects responsible for a difference between the up- and charm-quark sectors of the amplitude, including rescattering phases. Although power-suppressed they are calculable in QCD factorization. We then get a non-vanishing CP asymmetry already at ${\cal O}(\alpha_s^0)$. The effect for $B_s\to\gamma\gamma$ is again negligible whereas the direct CP asymmetry for $B_d\to\gamma\gamma$ is approximately $-10\%$. However, we do not expect these CP asymmetries to be measured in the near future because of the tiny branching ratios. We predict $B(\bar B_s\to\gamma\gamma) = (1.2^{+2.5}_{-0.7})\cdot 10^{-6}$ and $B(\bar B_d\to\gamma\gamma) = (3.1^{+6.7}_{-2.1})\cdot 10^{-8}$ which are both roughly two orders of magnitude below the current experimental upper limits. The large uncertainty in our prediction comes from the poorly known hadronic parameter $\lambda_B$.


\section{Conclusions}
We have presented a systematic and model-independent framework for the rare radiative decays $B\to V\gamma$ and $B\to\gamma\gamma$ based on the heavy-quark limit $m_b\gg\Lambda_{QCD}$. Quark-loop contributions are calculable in QCD factorization rather than being uncalculable long-distance effects. Non-factorizable long-distance corrections may still exist, but they are power-suppressed. Strong interaction phases from both hard-vertex and hard-spectator contributions are calculable and important for CP-violating observables. We have seen that weak-annihilation amplitudes in $B\to V\gamma$ are power-suppressed but numerically enhanced and calculable. We used them to estimate isospin breaking effects.

Our NLO predictions for the central values of the $B\to V\gamma$ branching ratios are $B(\bar B^0\!\to \bar K^{*0}\gamma) = (7.4^{+2.6}_{-2.4})\cdot 10^{-5}$ and $B(B^-\!\to\rho^-\gamma)  = (1.6^{+0.7}_{-0.5})\cdot 10^{-6}$. They are substantially larger than the leading logarithmic values when the same form factors are used. The dominant uncertainty comes from the variation of the nonperturbative input parameters, most notably from the $B\to V$ form factors. This situation, however, can be systematically improved. In particular, our approach allows for a consistent perturbative matching of the form factor to the short-distance part of the amplitude.

\section*{Acknowledgements}
I thank the organizers of QCD02 for inviting me to such an interesting conference and I am very grateful to Gerhard Buchalla for the extremely pleasant collaboration.

\end{document}